\documentclass[twoside]{dis09}
\usepackage[latin1]{inputenc}
\usepackage[dvips]{graphicx,epsfig,color}
\usepackage{wrapfig,rotating}
\usepackage{amssymb,amsmath,array}
\newcommand{\be}{\begin{eqnarray}}
\newcommand{\ee}{\end{eqnarray}}

\newcommand{\ket}[1]{\vert\,{#1}\rangle}

\begin{document}
\title{Chiral Odd and Chiral Even Generalized Parton Distributions in
Position Space}

\author{A. Mukherjee, R. Manohar$^1$ and D. Chakrabarti$^2$
%
%
\vspace{.3cm}\\
%
1-Department of Physics, Indian Institute of Technology Bombay\\
Powai, Mumbai 400076, India
%
\vspace{.1cm}\\
2-Department of Physics, Indian Institute of Technology Kanpur\\
Kanpur-208016, India
}

\maketitle

\begin{abstract}
We present the chiral odd generalized parton
distributions (GPDs) for non-zero skewness $\zeta$ in transverse and 
longitudinal position spaces by taking Fourier transform with respect 
to the transverse and longitudinal momentum transfer respectively using 
overlaps of light-front wave functions (LFWFs) in a simple model. We also
present the chiral even GPDs in positon space in a phenomenological model. 
\end{abstract}

\section{Introduction}

The chiral odd generalized transversity distribution $F_T$ is defined as
the off-forward matrix element of the bilocal tensor charge operator. It is
a leading twist generalized quark distribution.
It is parametrized in terms of four GPDs, namely $H_T$, $\tilde H_T$,
$E_T$ and $\tilde E_T$ in  the most general way \cite{markus,chiral,burchi}.
The chiral-odd GPDs give various informations on the transversely polarized
quark distribution in unpolarized and polarized nucleon. A relation for the
transverse total angular momentum of the quarks has been proposed
in \cite{burchi} which involves a
combination of second moments of $H_T, E_T$ and $\tilde H_T$ in the forward
limit. In a previous work we have investigated the chiral odd GPDs for 
a simple spin-$1/2$ composite particle for $\zeta=0$ in impact parameter 
space \cite{harleen}. Here, we investigate them for nonzero $\zeta$. 
Using an overlap formula in terms of the LFWFs both in the 
DGLAP ($n \rightarrow n$) and ERBL ( $n+1 \rightarrow n-1$) regions
\cite{ravi}, we investigate them in a simple model, namely  for
the quantum fluctuations  of a  lepton in QED at
one-loop order \cite{drell}. We generalize this analysis
by assigning a mass $M$ to the external electrons and a different
mass $m$ to the internal electron lines and a mass $\lambda$ to the
internal photon lines with $M < m + \lambda$ for stability.
In effect, we shall represent a spin-${1\over 2}$ system as a composite
of a spin-${1\over 2}$ fermion and a spin-$1$ vector boson
\cite{hadron_optics, dis,dip,dip2,marc}. This field theory inspired model
has the correct correlation between the Fock components of the state
as governed by the light-front eigenvalue equation, something that is 
extremely difficult to achieve in phenomenological models. GPDs in this 
model satisfy general properties like polynomiality and positivity. So it 
is interesting to investigate the general properties of GPDs in this model.
By taking Fourier transform (FT) with respect to $\Delta_\perp$, we express
the GPDs in transverse position space and by taking a FT with respect 
to $\zeta$ we express them in longitudinal position space. Qualitative 
LFWF for the meson and proton can be obtained by taking a derivative 
of the above LFWF with respect to the mass \cite{rad}. As a follow up, 
we present the distribution of partons in impact  parameter space in
a phenomenological model of the proton GPDs recently proposed in
\cite{simonetta}. 


\section{Generalized parton distributions in QED model}

We use the parametrization of \cite{burchi} for the chiral odd GPDs.
Following \cite{drell, hadron_optics}, we take a simple composite spin
$1/2$ state, namely an electron in QED at one loop to investigate the GPDs.
The light-front Fock state wavefunctions corresponding to the
quantum fluctuations of a physical electron can be systematically
evaluated in QED perturbation theory. The state is expanded in Fock
space and there are contributions from $\ket{e^- \gamma}$ and 
$\ket{e^- e^- e^+}$, in addition to renormalizing the one-electron state.
Both the two- and three-particle Fock state components are given 
in \cite{overlap}. In the domain $\zeta <x <1$, there are diagonal $2 \to 2$
overlaps. Using the overlap formula in \cite{ravi} we calculate the chiral  
odd GPDs in this kinematical region. In order to regulate the ultraviolet
divergences we use a cutoff $\Lambda$ on the transverse momentum
$k^\perp$. $\tilde H_T(x,\zeta,t)$ is zero in this model. 

We introduce the Fourier conjugate  $b_\perp$ (impact parameter)
of the transverse momentum transfer $\Delta_\perp$. The GPDs can
be expressed in impact parameter space by taking a Fourier transform
\cite{burchi}. In most experiments $\zeta$ is nonzero, and it is
of interest to investigate the chiral odd GPDs in $b_\perp$ space with 
nonzero $\zeta$. The probability interpretation of the impact parameter
dependent parton distributions for zero skewness is no longer possible 
as now
the transverse positions of the initial and final protons are different as  
there is a finite momentum transfer in the longitudinal direction. The GPDs 
in impact parameter space probe partons at transverse position $\mid b_\perp
\mid $ with the initial and final proton shifted by an amount of order
$\zeta \mid b_\perp \mid$. Note that this is independent of $x$ and even
when GPDs are integrated over $x$ in an amplitude, this information is still
there \cite{markus2}. Thus the chiral odd GPDs in impact parameter space
gives the spin orbit correlations of partons in protons with their centers
shifted with respect to each other. 
\begin{figure}{t}
\centering
\includegraphics[width=4.5cm,height=5cm,clip]{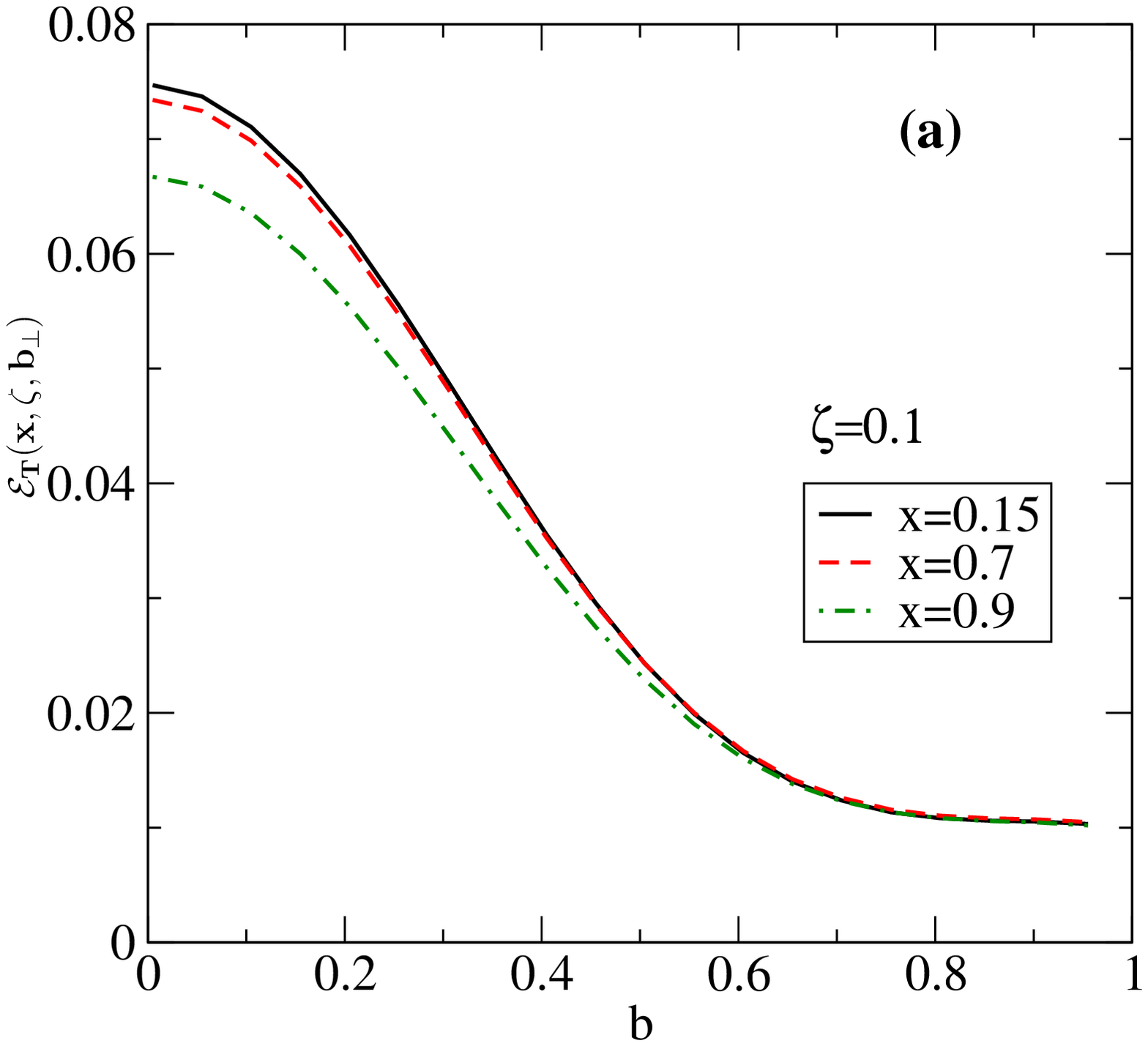}
\includegraphics[width=4.5cm,height=5cm,clip]{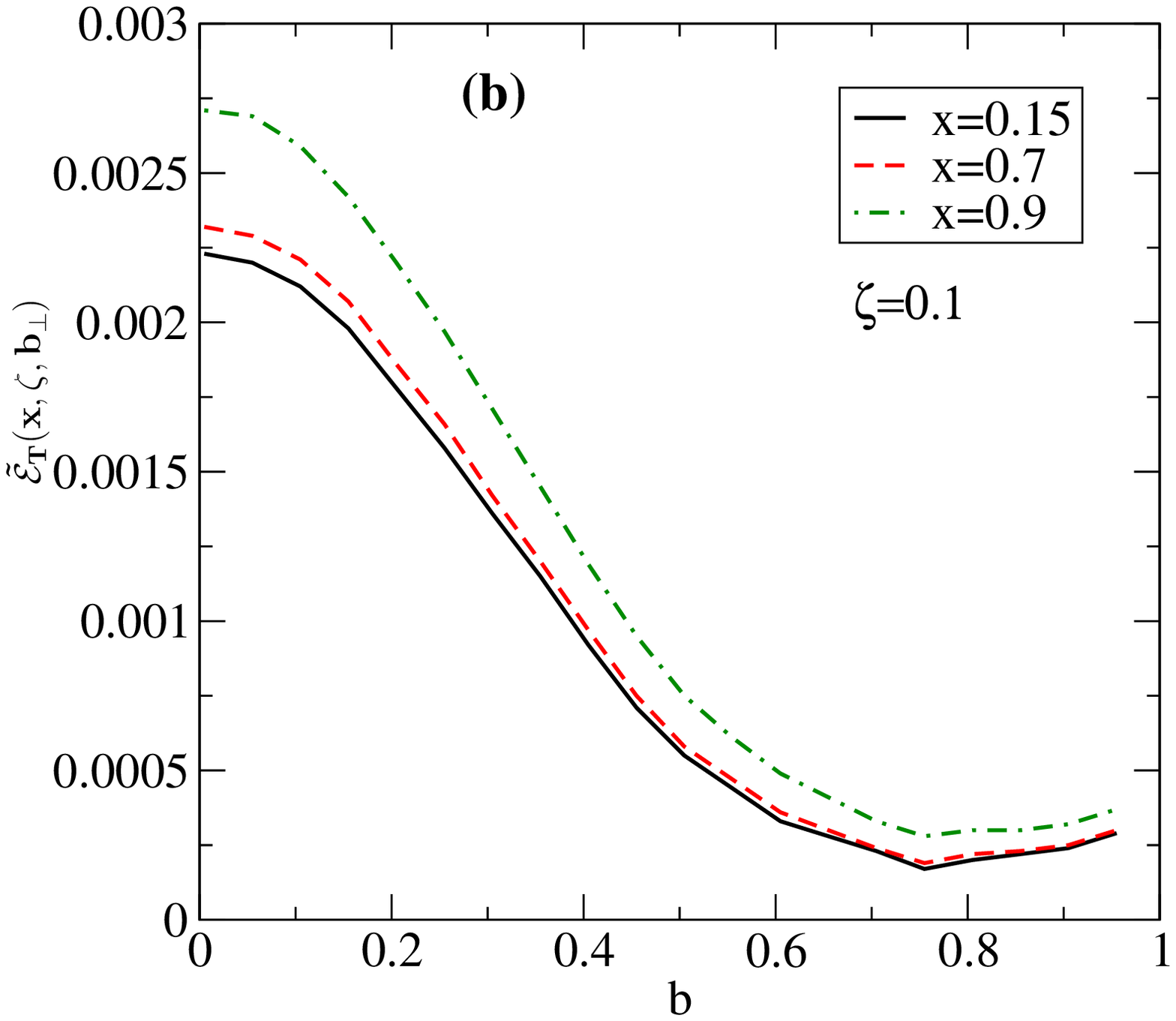}
\includegraphics[width=4.5cm,height=5cm,clip]{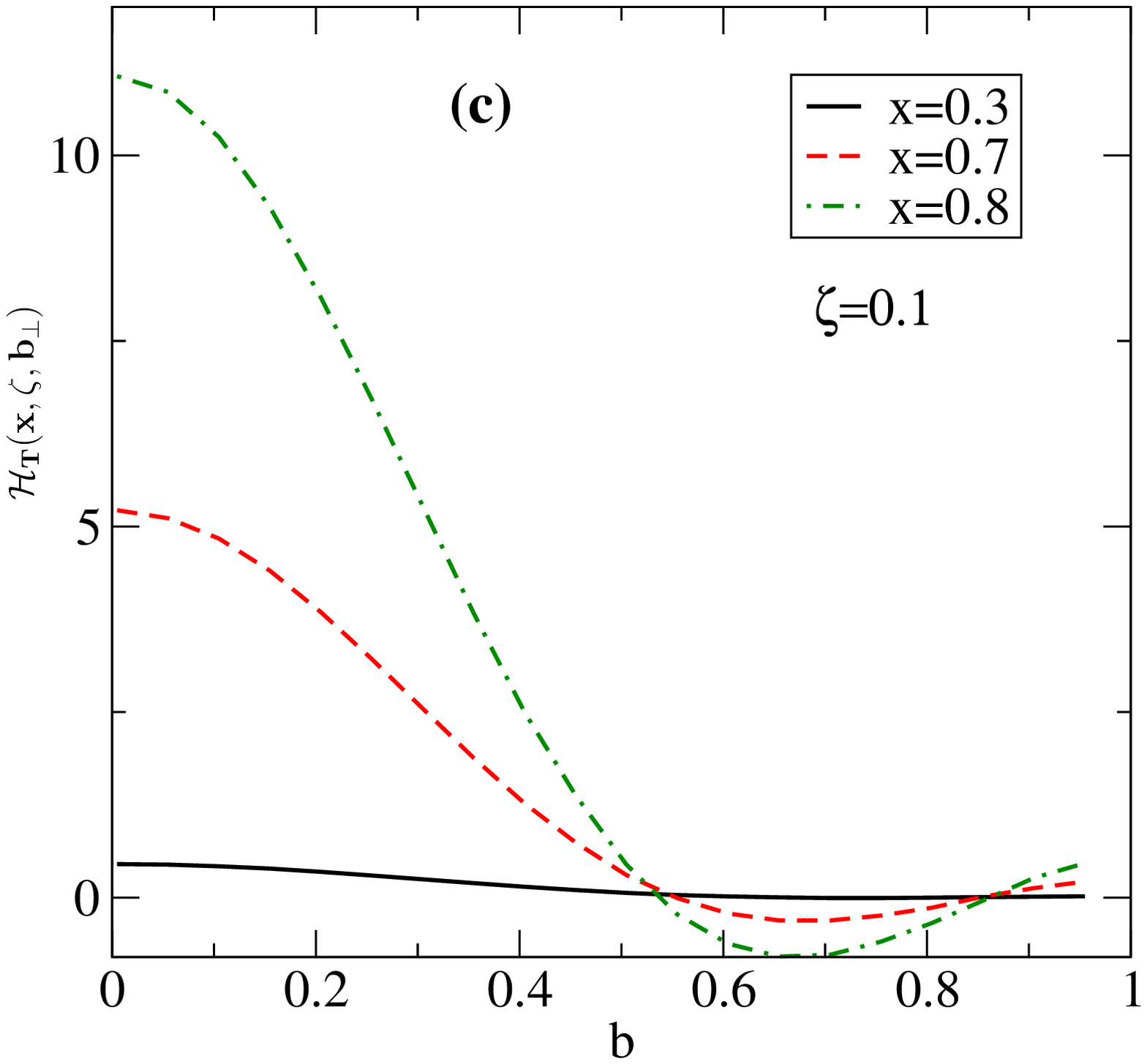}
\caption{\label{fig1} (Color online) Fourier spectrum of the chiral-odd
GPDs vs. $\mid b_\perp \mid $  for fixed $\zeta$  and
different values of $x$}
\end{figure}

Taking the Fourier transform (FT) with
respect to the transverse momentum transfer
$\Delta_\perp$ we get the GPDs in the transverse impact parameter space.
\be
{\cal E}_T(x,\zeta,b_\perp)&=&{1\over (2\pi)^2}\int d^2\Delta_\perp
e^{-i\Delta_\perp \cdot b_\perp} E_T(x,\zeta,t)\nonumber \\
&=&{1\over 2 \pi}\int \Delta d\Delta J_0(\Delta b) E_T(x,\zeta,t),
\ee
where $\Delta=|\Delta_\perp|$ and $b=|b_\perp|$. The other impact
parameter dependent GPDs $\tilde{\cal E}_T(x,\zeta,b_\perp)$ and 
${\cal H}_T(x,\zeta,b_\perp)$ can also be defined in the same way.
${\cal H}_T(x,\zeta,b_\perp)$ for a free Dirac particle is expected 
to be a delta function; the smearing in $\mid b_\perp \mid $ space 
is due to the spin correlation in the two-particle LFWFs.
Fig. \ref{fig1} shows the plots of the above three functions
for fixed $\zeta$ and different values of $x$. For given $\zeta$, the peak
of ${\cal H}_T(x,\zeta,b_\perp)$ as well as $\tilde 
{\cal E}_T(x,\zeta,b_\perp)$
increases with increase of $x$, however for ${\cal E}_T(x,\zeta,b_\perp)$
it decreases.

In \cite{wigner}, it was shown that the GPDs are related to a reduced Wigner
distribution $W_\Gamma(\vec{r},x)$  by a Fourier
transform.  For given $x$, this gives a 3D position space picture of the
partons inside the proton. If the probing wavelength is
comparable to or smaller than the Compton wavelength ${1\over M}$, where $M$
is the mass of the proton, electron-positron pairs will be created, as a
result, the static size of the system cannot be probed to a precision better
than  ${1\over M}$ in relativistic quantum theory. However, in light-front  
theory, transverse boosts are Galilean boosts which do not involve dynamics.
So one can still express the GPDs in transverse position or impact parameter
space and this picture is not spoilt by relativistic corrections. However,  
rotation involves dynamics here and rotational symmetry is lost. In
\cite{hadron_optics}, a longitudinal boost invariant impact parameter
$\sigma$ has been introduced which is conjugate to the
 longitudinal momentum transfer $\zeta$. It was shown that the DVCS
amplitude expressed in terms of the variables $\sigma, b_\perp$ show
diffraction
pattern analogous to diffractive scattering of a wave in optics where the
distribution in $\sigma$ measures the physical size of the scattering
center in a 1-D system. In analogy with optics, it was concluded that the
finite size of the $\zeta$ integration of the FT acts as a slit of finite
width and produces the diffraction pattern.

\begin{figure}
\centering
\includegraphics[width=4.5cm,height=5cm,clip]{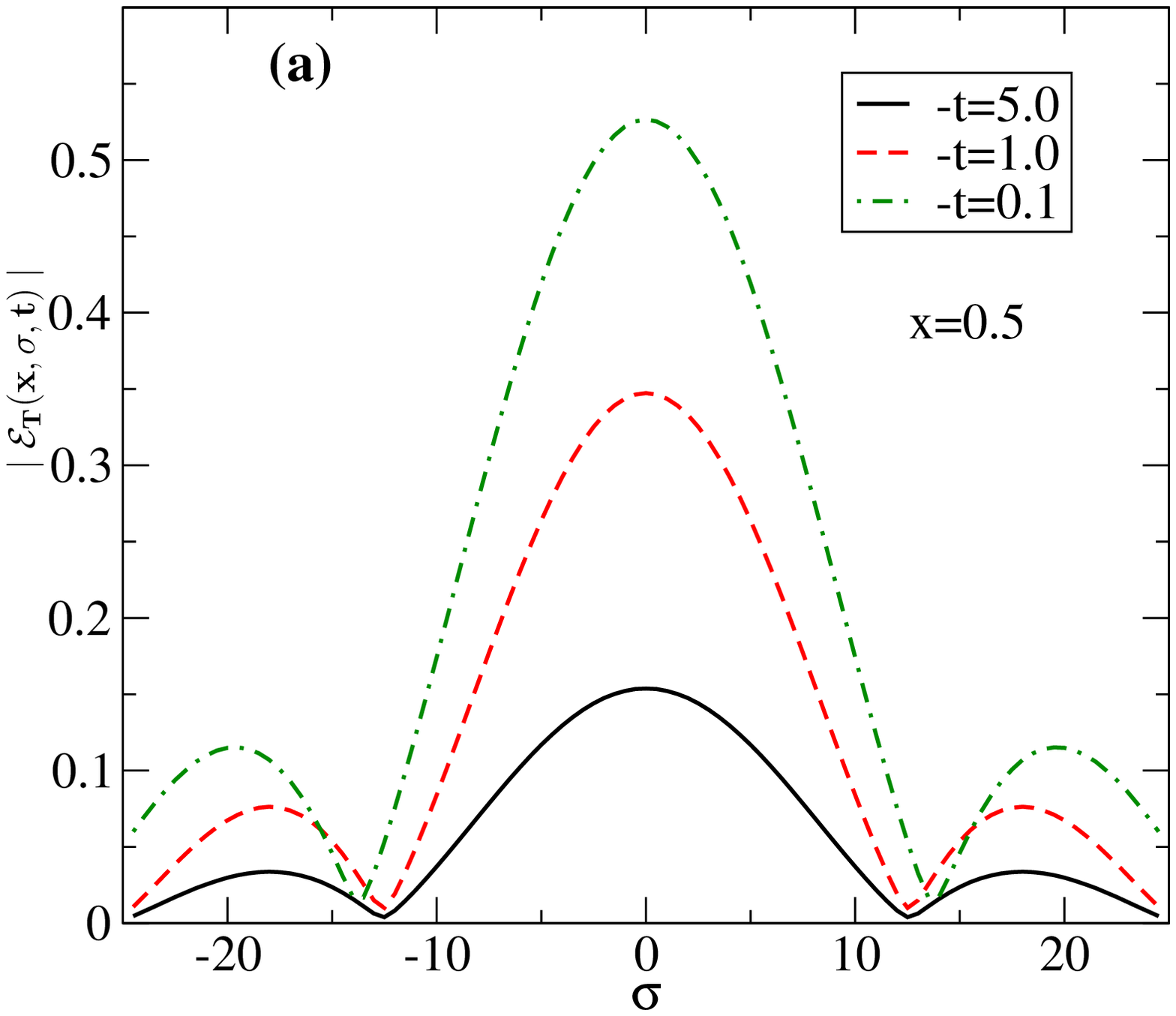}
\includegraphics[width=4.5cm,height=5cm,clip]{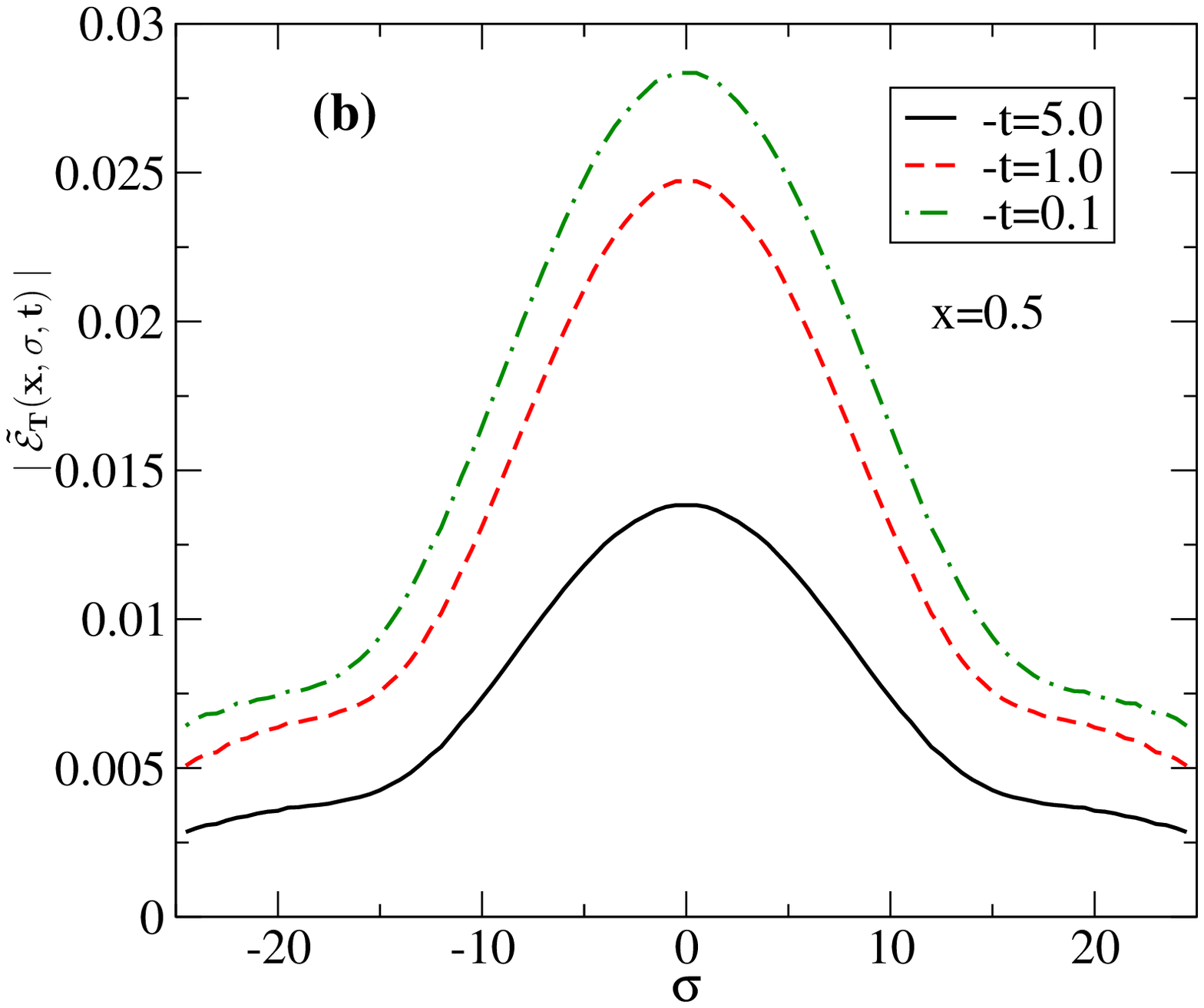}
\includegraphics[width=4.5cm,height=5cm,clip]{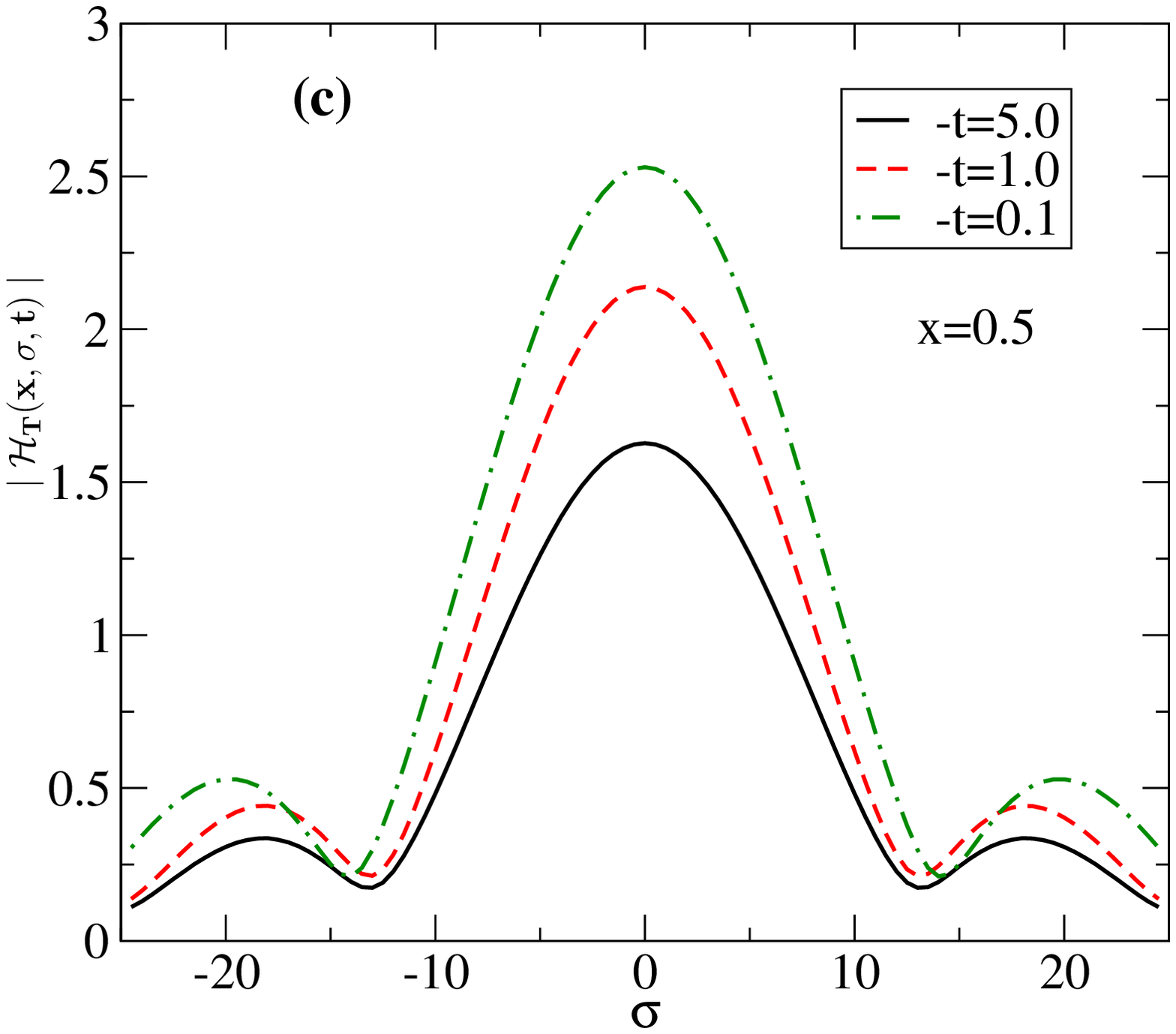}
\caption{\label{fig2} (Color online) Fourier spectrum of the chiral-odd
GPDs vs. $\sigma$  for fixed $x$  and different values of $-t$ in
${\mathrm{MeV}^2}$.}
\end{figure}
We define a boost invariant impact parameter conjugate to the  longitudinal
momentum transfer  as
$\sigma={1\over 2}b^-P^+$ \cite{hadron_optics}.  The chiral odd GPD $E_T$
in longitudinal position space is given by :
\be
{\cal E}_T(x,\sigma,t)&=& {1\over 2\pi}\int_0^{\zeta_f}d\zeta
e^{i{1\over 2}P^+\zeta b^-} E_T(x,\zeta,t)\nonumber \\
&=&{1\over 2 \pi}\int_0^{\zeta_f} d\zeta e^{i\sigma\zeta} E_T(x,\zeta,t).
\ee
Similarly one can obtain ${\cal H}_T(x,\sigma,t)$ and $\tilde {\cal E}_T
(x,\sigma,t)$ as well. Fig. \ref{fig2} shows the plots of the Fourier   
spectrum  of the chiral odd GPDs in
longitudinal position space  as a function of $\sigma$ for fixed $x=0.5$ and
different values of $t$. Both ${\cal E}_T(x,\sigma,t)$ and
${\cal H}_T(x,\sigma,t)$ show diffraction pattern as observed for the DVCS
amplitude in \cite{hadron_optics}; the minima occur at the sames values of
$\sigma$ in both cases.  However $\tilde {\cal E}_T(x,\sigma,t)$
does not show diffraction pattern.
This is due to the distinctively different behaviour of
$\tilde E_T(x,\zeta,t)$ with $\zeta$ compared to that of  $E_T(x,\zeta,t)$
and
$H_T(x,\zeta,t)$. $\tilde E_T(x,\zeta,t)$ rises smoothly from zero and
has no flat plateau in $\zeta$ and thus does not exhibit any diffraction
pattern when Fourier transformed with respect to $\zeta$.
The position of first minima in  Fig.\ref{fig2} is determined by $\zeta_f$.
For $-t=5.0$ and $1.0$, $\zeta_f \approx x=0.5$
 and thus the first minimum appears at the same position while for
$-t=0.1$,
$\zeta_f =\zeta_{max}\approx 0.45$ and the minimum appears  slightly
shifted.
This is analogous to the single slit optical diffraction pattern.
$\zeta_f$ here plays the role of the slit width.
Since the positions of the minima (measured from the center of
the diffraction pattern) are inversely proportional to the slit width,
the minima
move away from the center as the slit width (i.e., $\zeta_f$) decreases.
The  optical analogy of the diffraction pattern in $\sigma$ space has been
discussed in \cite{hadron_optics} in the context of DVCS amplitudes.

In ref. \cite{simonetta} a phenomenologically motivated parametrization of
the GPDs $H$ and $E$ are given by fitting the form factor data at zero
skewness. Regge type exchanges are incorporated. In Fig.3, we show the plots 
of the GPD $H$ for u-quarks in impact parameter space as a function of 
$b_\perp$ and also as a function of $x$.   

\begin{figure}
\centering
\includegraphics[width=5cm,height=5cm,clip]{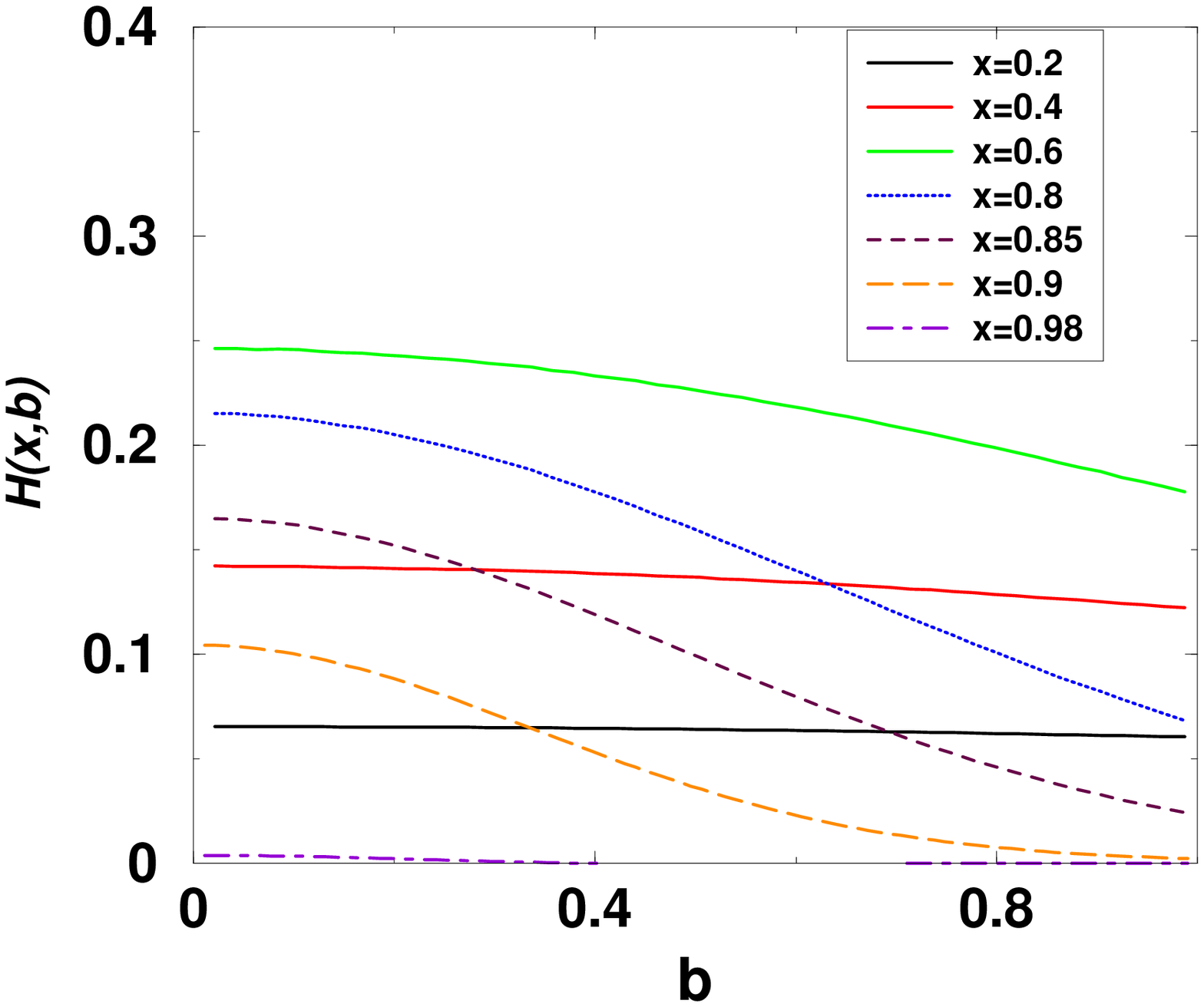}
\includegraphics[width=5cm,height=5cm,clip]{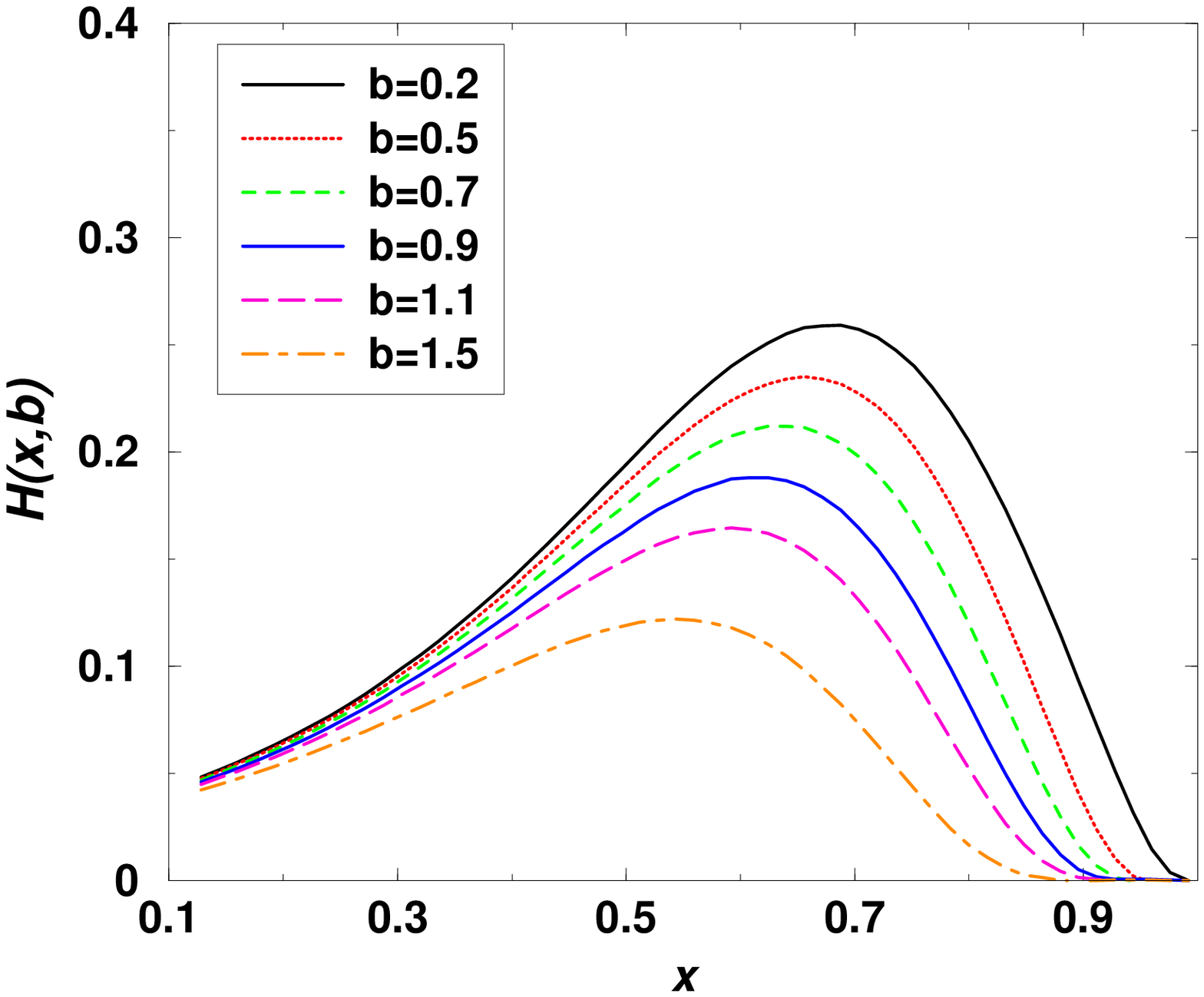}
\caption{\label{fig3} (Color online) Fourier spectrum of the chiral-even
GPDs for u-quarks in the proton in impact parameter space}
\end{figure}

\section{Conclusion}
In this work, we studied the chiral-odd GPDs in transverse and
longitudinal position space. Working in light-front gauge, we used
overlap formulas for the chiral odd GPDs in terms of proton
light-front wave functions  in the DGLAP region. 
We used a self consistent relativistic two-body model, namely the quantum
fluctuation  of an electron at one loop in QED. We used its most general 
form \cite{drell}, where we have a different mass for the external electron
and different masses for the internal electron and photon.
The impact parameter space representations are obtained by taking Fourier
transform of the GPDs with respect to the transverse momentum transfer.  
When $\zeta$ is non-zero, the initial and final proton are displaced in the
impact parameter space relative to each other by an amount proportional to 
$\zeta$. As this is the region probed by most experiments, it is of interest
to investigate this. By taking a Fourier transform with respect to $\zeta$  
we presented the GPDs in the boost invariant longitudinal position space    
variable $\sigma$. $H_T$ and $E_T$ show diffraction pattern in $\sigma$ space.
We also presented chiral even GPDs in impact parameter space using a
recently proposed parametrization.
\section{Acknowledgment}
AM thanks the organizers of DIS 2009 for invitation and support and IIT
Bombay for support.
\section{Bibliography}
 

\begin{footnotesize}


\end{footnotesize}



\begin{thebibliography}{99}
\bibitem{markus} M. Diehl, Eur. Phys. J. C {\bf 19}, 485 (2001).

\bibitem{chiral} M. Diehl and P. Hagler, Eur.Phys.J.{\bf C 44}, 87
(2005).

\bibitem{burchi} M. Burkardt, Phys. Rev. {\bf D 72}, 094020 (2005).

\bibitem{harleen} H. Dahiya, A. Mukherjee, Phys. Rev. {\bf D 77}, 045032
(2008).

\bibitem{ravi} D. Chakrabarti, R. Manohar, A. Mukherjee; Phys. Rev. {\bf D
79}, 034006 (2009). 

\bibitem{drell} S. J. Brodsky and S. D. Drell, Phys. Rev. {\bf D 22}, 2236
(1980).

\bibitem{hadron_optics}  S.~J.~Brodsky, D.~Chakrabarti, A.~Harindranath,
A.~Mukherjee and J.~P.~Vary, Phys.\ Lett.\  B {\bf 641}, 440 (2006); Phys.
Rev. {\bf D 75}, 014003 (2007).

\bibitem{dis} A. Harindranath, R. Kundu, W. M. Zhang, Phys.
Rev. {\bf D 59}, 094013 (1999);  A. Harindranath, A. Mukherjee,
R. Ratabole, Phys. Lett. {\bf B 476},  471 (2000); Phys. Rev.  
{\bf D 63}, 045006 (2001).

\bibitem{dip} D. Chakrabarti and A. Mukherjee, Phys. Rev. {\bf D 71}, 014038
(2005).

\bibitem{dip2} D. Chakrabarti, A. Mukherjee, Phys. Rev. {\bf D 72},
034013 (2005).

\bibitem{marc} A. Mukherjee and M. Vanderhaeghen, Phys. Lett. {\bf B 542},
245 Phys. Rev. {\bf D 67}, 085020 (2003).

\bibitem{rad} A. Mukherjee, I. V. Musatov, H. C. Pauli, A. V. Rayushkin, 
Phys. Rev. {\bf D 67}, 073014 (2003).  

\bibitem{overlap}  S. J. Brodsky, M. Diehl, D. S. Hwang, Nucl. Phys. {\bf B
596}, 99 (2001); M. Diehl, T. Feldmann, R. Jacob, P. Kroll, Nucl. Phys.
{\bf B 596}, 33 (2001), Erratum-ibid {\bf  605}, 647 (2001).

\bibitem{markus2} M. Diehl, Eur.Phys.J.{\bf C25},223-232,2002,
Erratum-ibid.{\bf C31}, 277-278,2003.

\bibitem{wigner} X. Ji, Phy. Rev. Lett. {\bf 91}, 062001 (2003);
A. Belitsky, X. Ji, F. Yuan, Phys. Rev. {\bf D 69} 074014
(2006).

\bibitem{simonetta} S. Ahmad, H. Honkanen, S. Liuti, S. Taneja , Phys. Rev.
{\bf D 75}, 094003 (2007).

\end{thebibliography}
\end{document}